\documentclass[aip,jap,preprint]{revtex4-1} 
\usepackage{amsmath, amssymb}
\usepackage{graphicx}
\usepackage{epstopdf}
\usepackage{dcolumn}
\usepackage{bm}
\usepackage{booktabs}

\begin{document}

\title{How to enable bulk-like martensitic transformation in epitaxial films}

\author{Marius Wodniok}
\author{Niclas Teichert}%
 \email{nteichert@physik.uni-bielefeld.de}
\author{Lars Helmich}%
\author{Andreas H\"utten}%
\affiliation{%
 Center for Spinelectronic Materials and Devices, Department of Physics, Bielefeld University, 33615 Bielefeld, Germany}

\date{\today}

\begin{abstract} 
The present study is dedicated to the influence of different substrate and buffer layer materials on the martensitic transformation in sputter deposited epitaxial shape memory Heusler alloys. For this, the magnetocaloric Heusler alloy Ni-Co-Mn-Al \cite{Teichert2015b} is grown on MgO(001), MgAl$_{2}$O$_{4}$(001), and MgO(001)/V substrates, which exhibit a lattice misfit to the Ni-Co-Mn-Al between $-1.2\%$ and $3.6\%$. By temperature dependent X-ray diffraction measurements it is shown that the optimum buffer layer for shape memory Heusler films is not one with minimum lattice misfit, but one with minimum Young's modulus and moderate misfit because an elastic buffer layer can deform during the martensitic transformation of the Heusler layer. Furthermore, epitaxial strain caused by a moderate lattice misfit does not significantly change the martensitic transformation temperatures. 
\end{abstract}

\pacs{Valid PACS appear here}
\keywords{Heusler, Magnetic Shape Memory Effect, Martensitic Transformation, Magnetocaloric Effect}
\maketitle


\section{\label{sec:Introduction}Introduction}
Magnetic shape memory Heusler alloys are promising materials for magnetic cooling and actuation applications due to their large inverse magnetocaloric effect\cite{Liu2012} and magnetic field induced strains\cite{Ullakko1996} related to the martensitic transformation (MT). Epitaxial thin films of these alloys are an ideal model system to study e.g. the martensitic microstructure \cite{Kaufmann2010,NiemannThesis} or interface effects \cite{Dutta2015}. However, a good model system should exhibit bulk-like behavior and as little substrate influence as possible. Usually, a single crystalline substrate leads to some amount of residual austenite at the interface and thus an incomplete MT in the epitaxial film. \cite{Teichert2015a}

In this work the influence of different underlayer materials on the MT of epitaxial Ni-Co-Mn-Al thin films has been investigated. The structure of Ni-Co-Mn-Al in the martensitic phase is \textit{L}1$_0$ and in the austenitic phase \textit{B}2. \cite{TeichertThesis}

The considered underlayers are single crystalline MgO(001) and MgAl$_2$O$_4$(001) substrates, and MgO(001)/V(35 nm) seed layer structures, which exhibit a lattice misfit $a_\text{substrate}/a_\text{film}-1$ between $-1.2\%$ and $3.6\%$.

\section{\label{sec:Experiment}Experimental Details}
Ni-Co-Mn-Al thin film samples were prepared by magnetron co-sputtering in an ultrahigh vacuum system with a base pressure of $<10^{-8}$\,mbar. During deposition the substrates were heated to $500^\circ$C and rotated at $10$\,rpm. On top of the Heusler film a $1$\,nm Si capping layer was deposited. 
In order to find the optimum substrate and seed layer for this shape memory Heusler compound, the following combinations were examined: MgO/Ni-Co-Mn-Al, MgAl$_2$O$_4$/Ni-Co-Mn-Al, and MgO/V/Ni-Co-Mn-Al. This choice is based on the low lattice mismatch below 4\% between the underlayer and the Heusler alloy. The substrate materials are MgO and MgAl$_2$O$_4$ single crystals with (001) surface. The thickness of the Ni-Co-Mn-Al films and the V seed layer were ascertained by X-ray reflectometry measurements to be $200$\,nm and 35\,nm, respectively.

The composition was determined to be Ni$_{38.7}$Co$_{8.4}$Mn$_{35.2}$Al$_{17.7}$ for all films by  X-ray fluorescence spectroscopy and energy dispersive X-ray spectroscopy measurements.

Structural characterization was done by X-ray diffraction (XRD) measurements in Bragg Brentano geometry using Cu K$_\alpha$ radiation. For the temperature dependent measurements a custom built LN$_2$ cryostat with a temperature range from 120\,K to 470\,K was used. In order to start these measurements in the austenite phase all temperature dependent measurements were started at 470\,K. 

\section{\label{sec:Results}Results and discussion}
\begin{figure}
\includegraphics[]{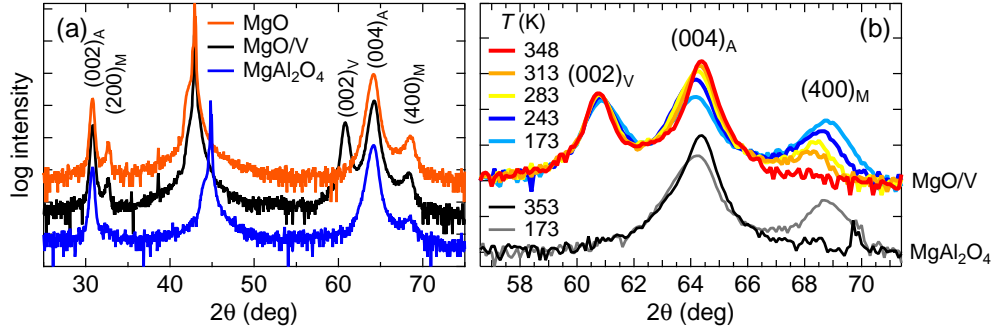}
\caption{XRD patterns. (a) room temperature XRD patterns of all investigated films, and (b) temperature dependence of the of the (002)$_\text V$, (004)$_\text A$, (400)$_\text M$ peaks.}
\label{Fig1}
\end{figure}

Figure \ref{Fig1}(a) shows the room temperature XRD patterns of the investigated films. Visible are the (002)$_\text{MgO}$ and (004)$_{\text{MgAl}_2\text O_4}$ and (002)$_\text V$ peaks at 42.9$^\circ$, 44.8$^\circ$, and 60.8$^\circ$, respectively. Further, the austenite peaks (002)$_\text A$ and (004)$_\text A$ are visible at 30.8$^\circ$ and 64.2$^\circ$. Martensite peaks (200)$_\text M$ and (400)$_\text M$ are found at 32.7$^\circ$ and 68.4$^\circ$. This indicates that all investigated films consist of a mixture of austenite and martensite at room temperature.

Using further measurements employing a 4-circle goniometer it was verified that the V layer grows epitaxially on the MgO and the Ni-Co-Mn-Al film grows epitaxially on all three underlayers with the same in-plane crystallographic orientation: The [100]$_\text A$ and [100]$_\text V$ direction are parallel to each other and to the [110] direction of the MgO or MgAl$_2$O$_4$ substrate.

\begin{figure}
\includegraphics[]{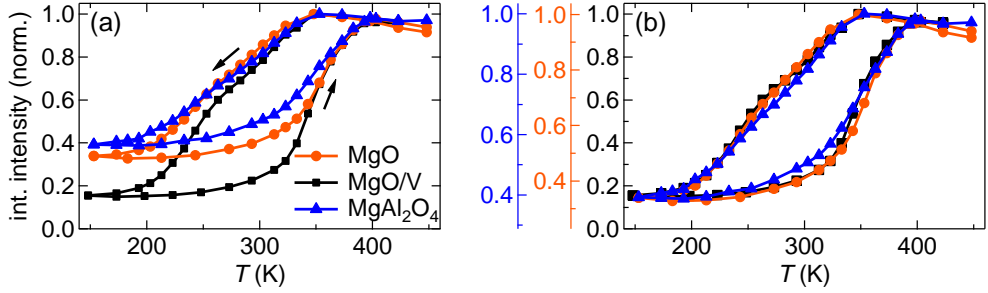}
\caption{Hysteresis of the MT. The temperature dependence of the integrated intensity of the (004)$_\text A$ peak is taken as a measure for the phase fraction of austenite. In (a) the the transformations are shown on the same y-axis and in (b) an overlay is generated by different y-axis for each film. The arrow indicates the direction of temperature change. The amount of residual austenite depends on the underlayer, but the temperature interval of the transformation does not.}
\label{Fig2}
\end{figure}

The temperature dependence of the (002)$_\text V$, (004)$_\text A$, (400)$_\text M$ peaks is shown in Fig.~\ref{Fig1}(b) for the MgO/V/Ni-Co-Mn-Al sample. At 470\,K the film is fully austenitic, which is seen from the absence of the (400)$_\text M$ reflection. This peak grows with decreasing temperature while the (004)$_\text A$ peak decreases. At 170\,K the integrated intensity of the (004)$_\text A$ peak is 15\% of the initial value at high temperature. The full temperature dependence of the integrated intensity of the (004)$_\text A$ peak is shown in Fig.~\ref{Fig2}(a) for all investigated films for heating and cooling. We take the intensity of the (004)$_\text A$ reflection as a measure for the austenite fraction of the film and conclude that for the film on MgO/V, a fraction of 15\% of the Ni-Co-Mn-Al film does not transform to martensite and remains at low temperature as residual austenite. For comparison, the film on MgAl$_2$O$_4$ exhibits 39\% of residual austenite and that one on MgO 33\%. It is therefore evident that the amount of residual austenite depends on the choice of the substrate.

From an overlay of the three hysteresis curves (cf. Fig.~\ref{Fig2}(b)) it is evident that the martensitic forward and reverse transformation occurs at approximately the same temperature for all samples, and hence, the choice of the substrate material does not significantly influence the transformation temperatures.

\begin{table}
\caption{\label{Tab1}Lattice constants, misfit, and Young's modulus of different substrates. The value for V is the average in-plane lattice constant as determined by XRD. Out-of-plane, 3.05\,\AA\ was found.}
\centering
\begin{tabular}[t]{lrrrr}
substrate &  $a_\text{substrate}$ (\AA) & $a_\text{Ni-Co-Mn-Al}$ (\AA) & misfit (\%) & $E$ (GPa) \\ 
\hline
MgO &  4.21 & 5.79 & 2.9 & 306\footnote{Reference [\onlinecite{Marklund1971}]} \\
MgO/V& 3.00 & 5.79 & 3.6 & 128\footnote{Reference [\onlinecite{Smithells1992}]}\\
MgAl$_2$O$_4$ & 8.08 & 5.79 & -1.2 & 275\footnote{Reference [\onlinecite{Askapour1993}]} \\
\end{tabular}
\end{table}

The lattice constants of the austenite, the substrates, and V seed layer are presented in Tab.~\ref{Tab1}. From this, it seems that a low lattice mismatch between austenite and underlayer is not preferred for a well performing shape memory thin film. Instead, the best results are obtained for the highest lattice mismatch of 3.6\%. 
Possibly, during film growth in the austenite state, the difference in lattice parameter is compensated by misfit dislocations, which can act as nucleation centers for martensitic nuclei close to the substrate and support a complete martensitic transformation.
However, it is also seen that the amount of residual austenite only slightly differs between MgAl$_2$O$_4$ and MgO where the difference in lattice constant is rather large, but strongly drops from MgO to V. So, it is assumed that the lattice mismatch is not the only influence on the transformation.

The habit planes in shape memory Heusler alloys are close to $\lbrace110\rbrace_\text A$ planes \cite{NiemannThesis} and thus, finite size (001) interfaces between martensite and the substrate are not possible without local deformations of the lattice. Therefore, it seems plausible that the elastic modulus of the substrate plays a role for the MT of the shape memory alloy close to the substrate.

\begin{figure}
\includegraphics[]{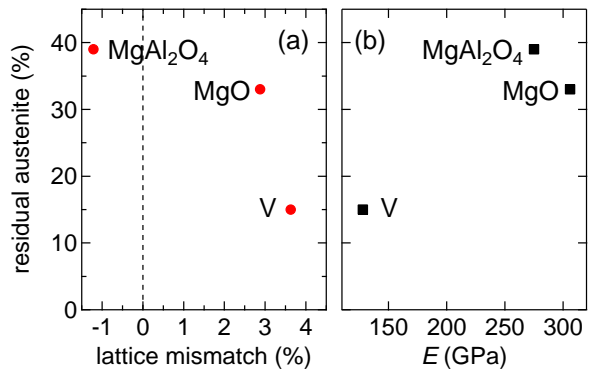}
\caption{Residual austenite in dependence of the lattice misfit (a) and Young's modulus (b). For a complete MT, small positive lattice misfit and an elastic underlayer are beneficial.}
\label{Fig3}
\end{figure}

To elucidate this, the dependence of the Young's modulus ($E$) of the substrates and seed layer on the amount of residual austenite is presented in Fig.~\ref{Fig3}. Since ascertaining the direction of the stress at the interface is difficult, the elastic constant for polycrystalline bulk is considered. Vanadium is by far the softest of the studied underlayer materials and allows the most complete MT. 
Furthermore, it is visible from Fig.\ref{Fig1}(b), that the (002)$_\text V$ peak broadens during the MT of the Heusler layer. It also decreases and shifts to higher scattering angles. 

\begin{figure}
\includegraphics[]{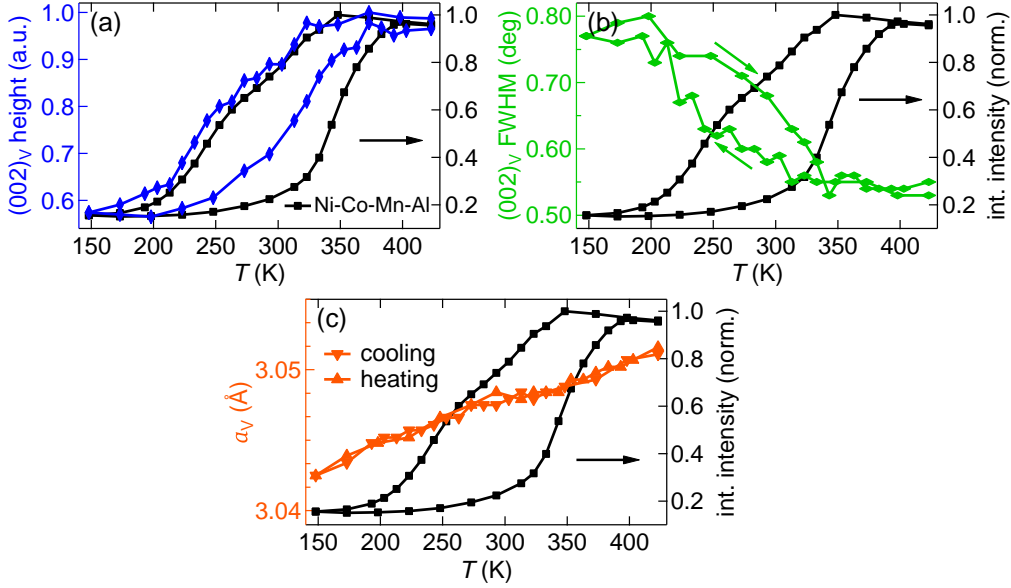}
\caption{Impact of the MT on the V buffer. The height (a) and FWHM (b) of the (002)$_\text V$ peak change significantly during the MT of the Ni-Co-Mn-Al film on top and exhibit a hysteresis. The black curves in all figure parts (right axis) show the MT of the Ni-Co-Mn-Al layer, see Fig.~\ref{Fig2}. The V out-of-plane lattice parameter (c) shows slight deviation from linearity and no hysteresis.} 
\label{Fig4}
\end{figure}

A detailed analysis of the (002)$_\text V$ peak is given in Fig.~\ref{Fig4}. Here, the peak height, FWHM, and the average out-of-plane lattice constant corresponding to the peak position are compared to the hysteresis of the MT of the Ni-Co-Mn-Al (black curve, right axis). 
The peak height (Fig.~\ref{Fig4}(a)) decreases and the FWHM (Fig.~\ref{Fig4}(b)) increases during the MT of the Heusler layer and both quantities exhibit a hysteresis. This peak broadening indicates that the coherent scattering length is reduced, possibly by microstrains, caused by the MT. 
It is striking that the change of the V peak on the cooling branch is nearly jointly with the forward MT, but on the heating branch, the reverse MT needs more overheating than the relaxation of the V buffer.
The reason for this is that the V layer does not undergo a phase transformation, but is elastically deformed. 
Upon heating, when the martensite becomes unstable, the restoring force leads to relaxation of the V, which probably causes some reverse transformation of the Ni-Co-Mn-Al close to the V layer, while the bulk of the Heusler film needs further overheating for a complete transformation.
However,the average out-of-plane lattice constant of the V buffer (Fig.~\ref{Fig4}(c)) shows only slight deviations from linearity during heating and cooling (thermal expansion) and no significant hysteresis. 
So, in order to elucidate the local changes of the V lattice constant caused by the MT of the Heusler layer, further measurements with high spacial resolution, employing e.g. temperature dependent high-resolution transmission electron microscopy, are needed.

Lastly, it is noted that we produced similar Ni-Co-Mn-Al films on MgO/Cr substrates (Cr: $a=2.88$\,\AA, $E=279$\,GPa\cite{Smithells1992}), with a lattice mismatch of only $-0.5\%$. In that film, no MT occurs due to interdiffusion between the Cr seed layer and the Ni-Co-Mn-Al film caused by the high deposition temperature. Deposition at lower temperature (300$^\circ$C) leads to a partial MT ($\approx 50\%$) but the results are not shown here because of impaired comparability to the other samples due to different preparation conditions.

\section{\label{sec:Conclusions}Conclusions}
In this paper, we investigated the the martensitic transformation of epitaxial Ni-Co-Mn-Al films in dependence of the underlayer elasticity and lattice misfit. We find that an underlayer with small positive lattice misfit to the austenite is preferred for a complete martensitic transformation. Furthermore, an elastic underlayer is strained during the MT in the film. The transformation temperatures are not significantly influenced by the substrate. Based on the presented results, V is identified as a good seed layer material for shape memory Heusler alloys.

\begin{acknowledgments}
The authors gratefully acknowledge funding by the DFG through SPP 1599 ``Ferroic Cooling''.
\end{acknowledgments}

\end{document}